\documentclass[aps,pre,twocolumn,showpacs,superscriptaddress]{revtex4}
\usepackage{amsmath}
\usepackage{graphicx}
\usepackage{pslatex}
\usepackage[T1]{fontenc}
\usepackage[latin1]{inputenc}

\begin{document}

\title{Searchability of networks}
\author{M. Rosvall}
\email{rosvall@tp.umu.se}
\affiliation{Department of Theoretical Physics, Ume{\aa} University,
901 87 Ume{\aa}, Sweden}
\affiliation{NORDITA, Blegdamsvej 17, Dk 2100,
Copenhagen, Denmark}\homepage{www.nordita.dk/research/complex}
\author{A. Gr{\"o}nlund}
\affiliation{Department of Theoretical Physics, Ume{\aa} University,
901 87 Ume{\aa}, Sweden}
\affiliation{NORDITA, Blegdamsvej 17, Dk 2100,
Copenhagen, Denmark}\homepage{www.nordita.dk/research/complex}
\author{P. Minnhagen}
\affiliation{Department of Theoretical Physics, Ume{\aa} University,
901 87 Ume{\aa}, Sweden}
\affiliation{NORDITA, Blegdamsvej 17, Dk 2100,
Copenhagen, Denmark}\homepage{www.nordita.dk/research/complex}
\author{K. Sneppen}
\affiliation{NORDITA, Blegdamsvej 17, Dk 2100,
Copenhagen, Denmark}\homepage{www.nordita.dk/research/complex}
\affiliation{The Niels Bohr Institute, Blegdamsvej 17, Dk 2100, Copenhagen, Denmark}

\date{\today}

\begin{abstract}
We investigate the searchability of complex systems in terms
of their interconnectedness.
Associating searchability with the number and size of branch points along
the paths between the nodes, we find that scale-free networks are 
relatively difficult to search, and thus that the abundance of scale-free
networks in nature and society may reflect an attempt to protect
local areas in a highly interconnected network from
nonrelated communication. In fact, starting from a random node, real-world networks with higher
order organization like modular or hierarchical structure are
even more difficult to navigate than random scale-free networks.
The searchability at the node level opens the possibility for a generalized hierarchy
measure that captures both the hierarchy in the usual terms of trees as in military structures,
and the intrinsic hierarchical nature of 
topological hierarchies for scale-free networks as in the Internet.
\end{abstract}
\pacs{89.75.Fb, 89.70.+c}
\maketitle

\section{Introduction}
\begin{figure}[!htbp]
\includegraphics[width=\columnwidth]{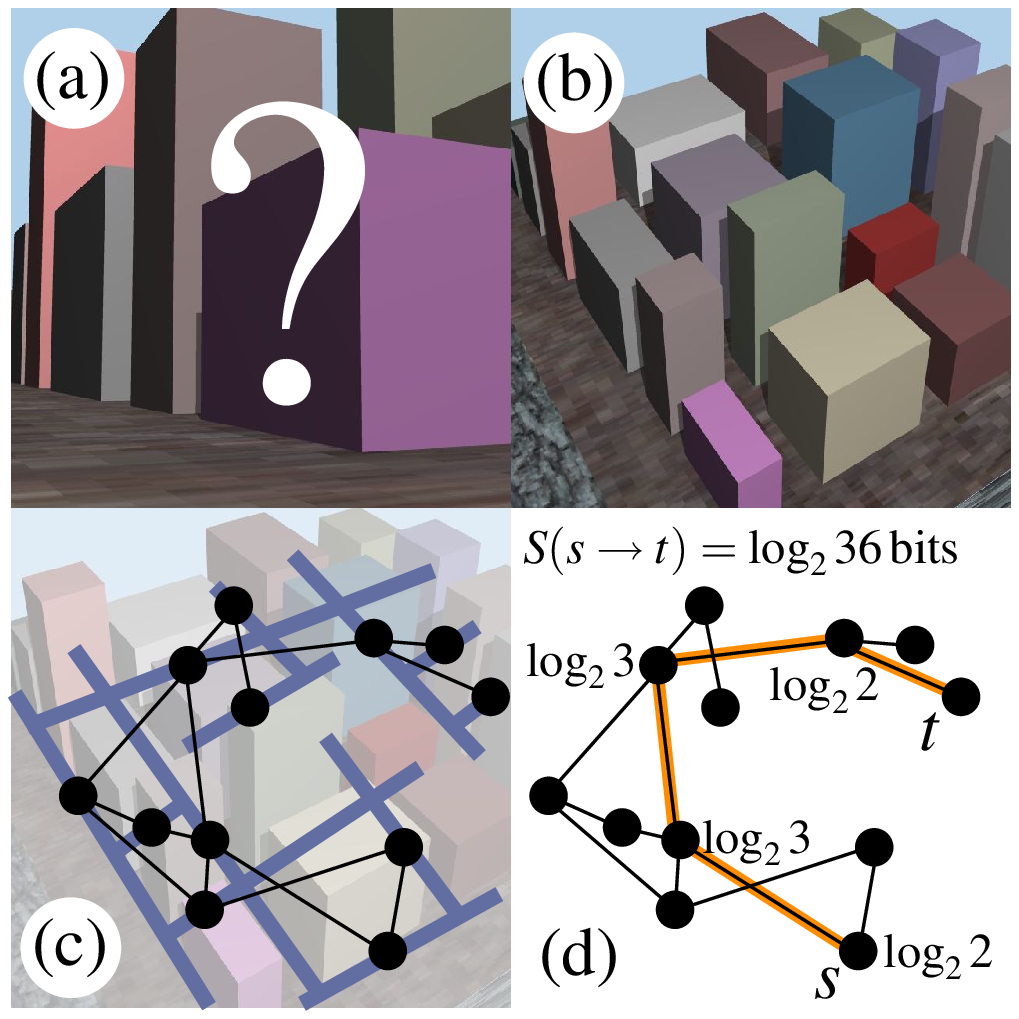}
\caption[] {(Color online) An example where the search information is an important concept. 
(a) illustrates a visitor's 
perspective of an unknown city. The visitor therefore asks
a citizen with the perspective (b) of the city, or rather the
higher abstraction level (c). This level is the dual map of the 
city, a network where streets are identified as nodes and
intersections between streets as links between the nodes.
We use this level to quantify the search information in (d); the
average number of yes-no questions the visitor must ask the
citizen to find a specific street. The necessary information to 
walk the shortest path from $s$ in the
lower right corner to $t$ in the upper right corner is $\log_2 36$ bits or roughly 6 yes/no-questions.
}
\label{fig1}
\end{figure}

Each element interacts directly only with a
few particular elements in most complex systems. 
Distant parts of the network thereby formed 
can consequently communicate through sequences of local
interactions. In this way all parts of the network can
be reached from other parts, but not all such communications are
equally easy or accurate \cite{friedkin,kochen,milgram,rosvall}.
The purpose of this paper is to investigate the interplay between searchability of a network
and the network structure. By searchability or navigability we mean the difficulty of sending a signal between 
two nodes in a network without disturbing the remaining network.
We use a city-street network to
illustrate the concept of navigability in networks \cite{city}.
As in Fig.\ \ref{fig1}(c) the streets are identified
as nodes and intersections between the streets as links between the nodes.
From this point of view, the above statement reads:
A pedestrian or driver on a street in a city, can by multiple choices
reach any other street in the city via the intersections.
However, not all streets are as easy to find, and the difficulty
of finding a street may vary from city to city. 

In the current paper we investigate how
different network topologies influence the average amount of
information that is needed to send a signal from one node to another node
in the network. We consistently concentrate on specific signaling, 
and focus only on locating one specific node
without disturbing the remaining network.
This is different from the nonspecific
broadcasting where any input is amplified by all exit
links of every node along all paths like in
spreading of spam or propagation of diseases and computer viruses \cite{moreno,newmanVirus}. 
We present a quantification of the specific signaling
and justify our choice of measure by
its minimum information property.

\section{Search information}
\begin{figure}[htbp]
\includegraphics[width=\columnwidth]{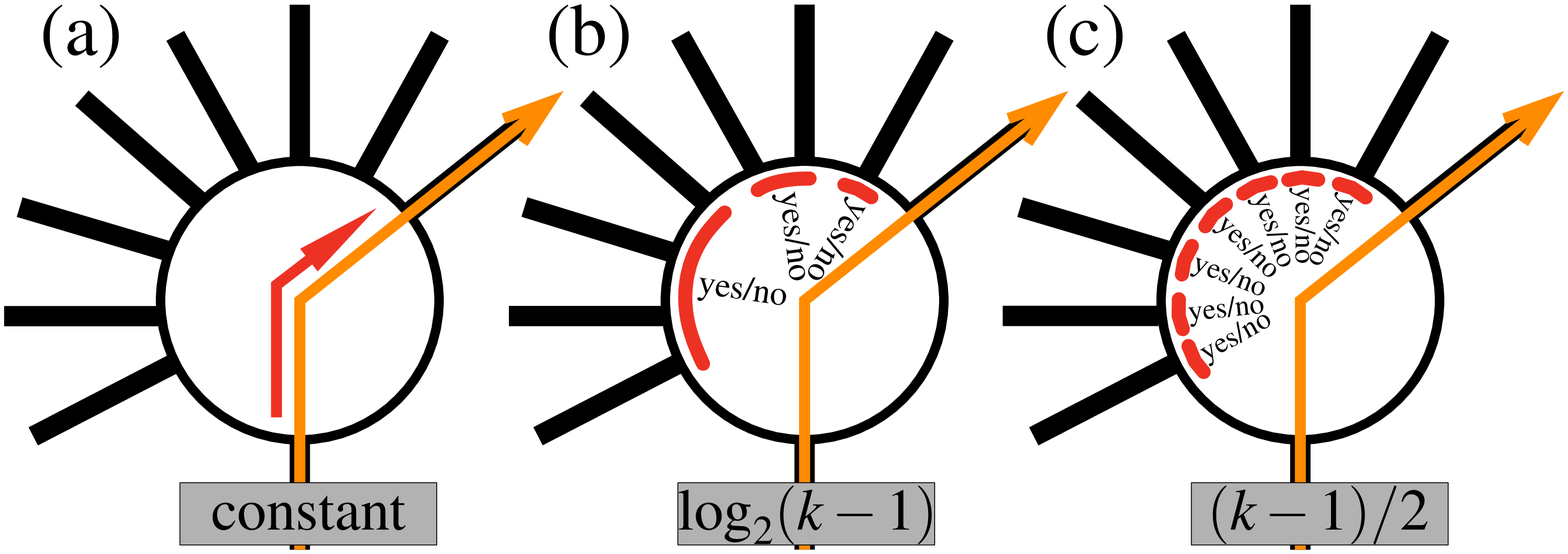}
\caption{\label{fig1b}%
(Color online) The information cost at each node depend on the ordering of the link. 
(a) The information cost does not depend on the degree if there is only one
possible link.
(b) It is possible to ask yes-no questions and
successively eliminate groups of wrong exits if the links are ordered.
Every yes-no question optimally reduces the number of possible links with $1/2$ and
the cost is $\log_2(k-1)$ to find the correct exit link.
(c) If the links are unordered such groupings are impossible and every exit link must be considered. 
In such a scenario the average number of necessary yes-no questions is $(k-1)/2$.
}
\end{figure}

We consider a specific signal, or a walker, on a network and
assume that the specific signal
from a source $s$ to a target $t$
is a signal that travels along the shortest path,
and thereby minimizes the disturbance on other nodes.
This assumption is made on the basis that the shortest path
is a good estimate for typical traffic in a network \cite{krioukov}.
We will later discuss the alternative model,
to follow the minimal information path, not necessarily
coinciding with the shortest path. 
The minimal amount of information needed to follow a
specific shortest path is determined by the degrees
of the nodes along the path, i.e., the number and size
of the branch points between the nodes.
That is, a walker on the network first has to choose
the right exit link (we call it the path link)
among the $k_s$ possible links from $s$.
The cost depends on the available global information
on the node and the way the information is organized at the node \cite{adamic,adilson}. 
If no information is available, the choice must be random, and the walker
will perform a random walk. We here consider scenarios where the network
represents the communication backbone of a system with available information
on the node level, e.g., social networks \cite{milgram,rosvall}, computer networks \cite{adilson}, city networks \cite{jiang,city}, etc.
In principle, if the exit links are unordered, one 
yes-no question must be asked for every exit link
to find the path link. On average this would give rise to an
average cost of $k/2$ yes-no questions or $(k-1)/2$ if the
arrival link at the node is known and one link immediately can
be excluded. This is illustrated in Fig.\ \ref{fig1b}(c).

The other extreme situation is when the exit path somehow is
given by default or the information cost can be neglected in comparison to
the walk on the shortest path itself [Fig.\ \ref{fig1b}(a)]. 
We here focus on the case where the links are ordered, 
like intersections along a road. In this case a question can be used
to reduce the possible outcomes by a factor $2$.
A city example: The yes-no answer to ``Does anyone of the eight closest roads
lead toward the station?'' reduces the outcome to eight roads if there were 16 possible intersecting roads
to choose from.
The total number of bits, or roughly the number of yes-no questions, necessary to find the path link 
is $\log_2(k)$ or $\log_2(k-1)$ if the arrival node is known to not lead to the target as in Fig.\ \ref{fig1b}(b).

\begin{figure*}[tpbh]
\includegraphics[width=\textwidth]{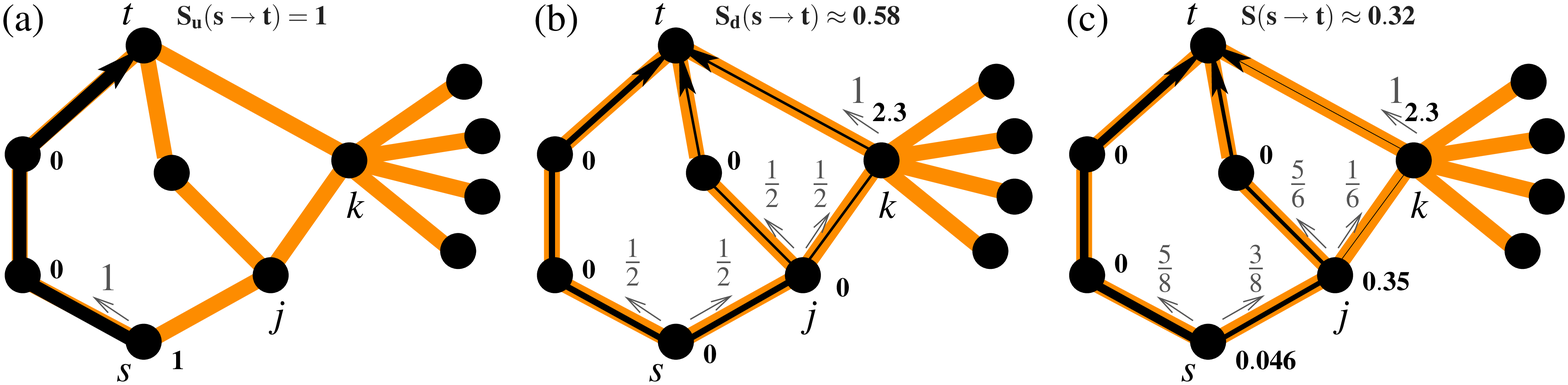}
\caption[] {%
(Color online) Search information with degenerate paths between the source $s$ and target $t$.
The numbers around the nodes indicate with what probability the link is chosen on the walk to $t$.
The boldfaced number is the information cost in bits at the given node.
The total information cost $S(s \to t)$ above every network is given by the average cost over
all paths marked with black lines (the sum of the costs at the nodes along the paths) weighted 
with the probability to walk the path (width of black lines).
We present three scenarios.
(a) The walker aims to take a specific shortest path in, the cheapest informationwise.
(b) The walker chooses between two exits, both on the shortest path, randomly.
This results in a lower total information cost since a random choice does not cost any information,
even though some walks will go through the expensive hub to the right (2.3 bits).
(c) The walker chooses to minimize the average information cost between $s$ and $t$.
The difference between (b) and (c) is clear from the choice at node $j$.
In (c) more information is used at this node to avoid the 
higher cost of going to the hub to the right.
This is completely avoided in (a) by going to the left at $s$, but at a higher information cost.
}
\label{fig2}
\end{figure*}

That is, $\log_2(k_s)$ bits of information are necessary at the start node $s$, 
where $k_s$ is the degree of $s$.
Subsequently the walker at each node $j\in p(s,t)$ along the path
$p(s,t)$ has to choose the particular exit link
along the path. Given the knowledge to follow the path
to $j$, there are $k_j-1$ unknown exit links from $j$, 
and the information needed to make the next step is $\log_2(k_j-1)$.
As a result the total information needed to follow the path is
\begin{equation}
S_u(p(s,t)) \;=\; \log_2(k_s)+ \sum_{j\in p(s,t)} \log_2(k_j-1),
\label{single-path}
\end{equation}
where $p(s,t)$ includes nodes on the path between $s$ and $t$, but not
the start and end nodes $s$ and $t$ [see Fig.\ \ref{fig2}(a)].
We use the notation $S_u$ to emphasize that the walk is a result 
of decisions for a specific and unique path and repeat that we use $\log_2(k_j-1)$ 
at every step but the first since the link of arrival is known (Fig.\ \ref{fig1b}).

If there is more than one shortest path between $s$ and $t$
the information needed to travel along one of the shortest paths
has to include the thereby added degenerate possibilities \cite{friedkin}.
Degenerate paths imply that more than one exit link can lead
the walker closer to the target from each node, and should be reflected in
a decreased path information $S_d(\{ p(s,t)\})$;
the subscript $d$ is for degenerate paths and $\{ p(s,t)\}$ is for the set of paths
between $s$ and $t$.
If a node $j$ has $k_j$ links, of which $\eta_j$ links
point toward the target node $t$, then the
number of bits to locate one of the correct exits is
reduced to $\log_2[(k_j-1)/\eta_j]$
[and to $\log_2(k_s/\eta_s)$ for the first step at the source node $s$].
In this definition we make the assumption that the probability of choosing 
any exit link on a shortest path from the current node is equal.
Therefore each of the degenerate paths
will be selected with a different probability,
as indicated in the example of Fig.\ \ref{fig2}(b).
That is, each path in the set of degenerate paths $\{p(s,t)\}$ is selected with
probability
\begin{equation}
P(p(s,t)) \;=\; \frac{1}{\eta_s} \prod_{j\in p(s,t)} \frac{1}{\eta_j}.
\label{branching}
\end{equation}
The average number of bits needed to follow a random
shortest path is accordingly
\begin{equation}
S_d(s \to t) \; = \; \sum_{\{p(s,t)\}} P(p(s,t)) \cdot \log_2 \left(
\frac{k_s}{\eta_s} \; \prod_j \frac{k_j-1}{\eta_j} \right) \;.
\end{equation}
This simplifies to Eq.\ (\ref{single-path}) in the case where there is
only one degenerate path.
When there are degenerate paths between $s$ and $t$, 
$S_d(s \to t)$ does not distinguish paths that are difficult
to follow from the easier ones, but just averages.

The average path information $S_d(s \to t)$ is closely
related to the earlier introduced search information
\cite{pnas,city,horizon,bit}:
\begin{equation}
S(s \to t) =
-\log_2 \left(
\sum_{\{p(s,t)\}}\frac{1}{k_s}\prod_{j}\frac{1}{k_j - 1} \right),
\label{Smin}
\end{equation}
where the sum runs over the set $\{p(s,t)\}$ of degenerate shortest
paths between $s$ and $t$. Thus, again, if there are no degenerate shortest
paths, $S(s \to t) = S_u(s \to t)$. If there are
degenerate paths, the relative weighting of these paths differs. In
the $S_d$ measure each path is weighted according to the branching
of shortest path shown in Fig.\ \ref{fig2}(b), and is thus the typical
information needed to follow a random branch of one of the
shortest paths through the network. In contrast $S$ measures the
minimal information value of knowing the full path and 
the subscript $m$ for minimal is omitted.
$S$ is defined as $-\log_2$ of
the probability that a nonguided signal emitted from $s$ arrives at
$t$ with minimal number of steps. For all networks we have tested $S_d$
is maximally a few percent larger than $S$, reflecting that situations where
one of the branches is substantially more difficult to travel only
gives a small additional correction to $S_d$ (see Fig. \ref{fig3}). 
Also, we always
found indistinguishable results when we analyzed the networks in terms
of the conditional uniform test
$S-S(\mathrm{random})$ or in terms of $S_d-S_d(\mathrm{random})$ \cite{maslov2002,maslovInternet}.

To get the corresponding probabilities to follow a given path as in 
Eq.\ (\ref{branching}) we present a simple example of the minimum information property
of $S$, and choose the path from $j$ to $t$ in Fig.\ \ref{fig2}(c) as an example path. Let the probability
to take the left path be $q_1$ and the right path via the hub be $q_2=1-q_1$ and
further the probability to reach the target be $p_1$ after the left choice is taken and
$p_2$ if the right choice is taken. 
The probability to choose the link down to the left is 0, since it is not on a shortest path to $t$.
Then the total information cost from $j$ to $t$
is
\begin{eqnarray}\label{Sderive}
S(j \to t) = ( \log_2{3}+q_1\log_2{q_1}+q_2\log_2{q_2} )
-(q_1\log_2{p_1}+q_2\log_2{p_2}),\nonumber
\end{eqnarray} 
where the first parantheses on the right-hand side is the information cost to pay at node $j$.
The full expression of this term reads 
\begin{eqnarray}
\sum_{i=1}^{3}-\frac{1}{3}\log_2(\frac{1}{3})-\sum_{i=1}^{3}-q_i\log_2(q_i),
\end{eqnarray}
and is the difference between the information entropy of a random choice and
the information entropy of the actual choice---the meaningful information of the choice.
The information cost payed at node $j$ ensures that the walker takes the path to the left with probability $q_1$ and
to the right with probability $q_2$. This is equivalent to the meaningful information content of
a policeman in the crossing who 
points toward the left with probability $q_1$, 
to the right with probability $q_2$, and never down to the left (since it is not on a shortest path to $t$).
The remaining two terms in Eq.\ (\ref{Sderive}) represents the cost from the next step to the target as 
two contributions according to Eq.\ (\ref{Smin}), weighted with the probabilities $q_1$ and $q_2$ of
choosing the paths. 

We set $\mathrm{d}S/\mathrm{d}q_1=0$ to find the minimum. With $q_2=1-q_1$ we get
\begin{equation}
0=\log_2{q_1}-\log_2(1-q_1)-\log_2{p_1}+\log_2{p_2},
\end{equation}  
or $q_1/q_2=p_1/p_2$, satisfied by 
\begin{eqnarray}
q_1=p_1/(p_1+p_2),\:q_2=p_2/(p_1+p_2).
\end{eqnarray} 
Inserting this back in Eq.\ (\ref{Sderive}) gives
\begin{eqnarray}
S(j \to t) =-\log_2\left(\frac{1}{3}p_1+\frac{1}{3}p_2\right),
\end{eqnarray} 
which is identical to Eq.\ (\ref{Smin}). 
Effectively $q_1$ and $q_2$ weight the probability of choosing an exit from $j$
with the difficulty of following it. For example, paths that contain
large hubs will be suppressed because the probability of following such paths randomly is lower.

To be able to characterize the complete network in terms of
searchability we define $S$ as the average of pairwise search information between nodes over all pairs of nodes
\begin{align}
S = \frac{1}{N(N-1)}\sum_{s \ne t}S(s \to t).
\end{align}
Thus, although $S$ is defined in terms of global random walkers
it should be interpreted as subsequent and local minimization of information costs
to navigate to a target node. 
Thus, it is different from the random walker approach 
that has been used to 
characterize topological features of networks \cite{bilke,monasson}, 
including first passage times \cite{noh},
large scale modular features \cite{eriksen}, and
search utilizing topological features \cite{adamic}.
Neither should the search information with its logarithm of base 2 be mixed up with
entropy measures associated with the degree distribution \cite{sole},
measures related to the dominating eigenvector
of the adjacency matrix \cite{demetrius}, or
different flows on networks like betweenness centrality and closeness centrality \cite{freeman,between,borgatti}.
Instead $S$ measures the amount of information that turns a random walker to
a directed walker that follows a shortest path (or any other chosen path) between the source $s$ and target $t$.

Some insight into the search information $S$, which also makes the
difference from a pure entropy measure clear, is obtained if we consider the
simple average along one of the shortest paths, and ignore
information associated with having arrived from a link that
cannot be leading closer to $t$:
\begin{eqnarray}
{\cal S}(s,t) \; & = &\; \sum_{p(s,t)} P(p(s,t)) \cdot
\log_2 \left( k_s \; \prod_j k_j \right)
\end{eqnarray}
with a total average path information
\begin{eqnarray}
{\cal S}\; & = & \; \frac{1}{N(N-1)}\sum_{j=1}^N  b(j) \log_2 ( k_j),
\label{simple}
\end{eqnarray}
which differs from a pure entropy measure of the form $\sum p \log p$ since
$b(j)$ is proportional to $k_j$ only when the walk is random.
Here $b(j)$ is the traffic betweenness of the node $j$,
defined as the number of shortest paths between pairs of nodes
in the network that pass through node $j$,
including paths that start at $j$ or paths that end at $j$.
This traffic betweenness differs from the usual betweenness
\cite{freeman,between} by the different treatment of degenerate paths,
in the sense that a given degenerate
path contributes to betweenness with a weight
given by the difficulty of walking the path according to Eq.\ (\ref{Smin}).
In practice, in all the real networks that we have investigated, 
we found that the difference is negligible.
We thus expect relatively large $S$ values for networks
(1) where there are many nodes on the shortest path
between other nodes [most $b(i)$ large], and (2) where most traffic goes
through highly connected nodes.
Point 1 predicts large $S$ for modular networks,
whereas point 2 suggests relatively large $S$ for
networks with broad degree distributions.
The path length is indirectly coupled to points 1 and 2;
stringy networks as well as regular networks with
long average path lengths have high $S$ and
starlike networks have small $S$ despite point 2,
because of the very short paths.
In the remaining part of this paper we will
examine the interplay between $S$ and
global topology in detail.

\section{Search information in model networks}
The search information is topology dependent, and in this section we present
how $S$ captures the average degree, the degree distribution, and higher order topological organization 
of the networks.

\begin{figure}[htbp]
\includegraphics[width=\columnwidth]{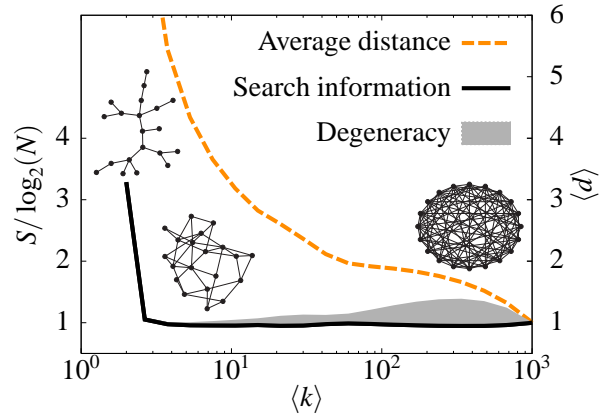}
\caption[] {(Color online) Search information of random Erd{\H o}s-Rényi (ER) networks as a function
of average degree $\langle k \rangle$. The number of nodes is $N=10^3$ and we 
keep the networks connected.
The shaded area is the contribution from degenerate paths with
the upper edge corresponding to the definition of $S_u$ in Fig.\ \ref{fig2}(a).
The definition in Fig.\ \ref{fig2}(b) is inseparable in this plot from the search information
according to Fig.\ \ref{fig2}(c).
The degenerate paths make the highly connected networks more searchable, 
mainly due to degenerate paths of length 2 between each pair of nodes.
Thus the resulting $S$ is lower than that obtained when considering
information associated to locating just one of the shortest paths
(upper border of shaded area). For very low degrees,
$\langle k \rangle \sim 2$, the organization of the networks
opens for a broad range of different topologies with very
different searchability; the average shortest path increases and
finally no degenerate paths exist.}
\label{fig3}
\end{figure}

Figure \ref{fig3} shows how the $S$ depends on average degree
$\langle k \rangle$ in a random network.
The lower curve is the total $S$ and the shaded area represents
the contribution from degenerate paths.
The upper border of the shaded area is consequently $S_u$, the search information
without degenerate paths.
$S_d$, ($S_u \ge S_d \ge S$), that weights paths according to branch points
along the paths, is within the shaded area (although it is indistinguishable from the lower curve in the present case). 
Notice that the figure mostly examines very high
$\langle k\rangle$ values where most pairs of
nodes are connected by multiple degenerate paths of length 2.
This explains the reduction in search information due to degenerate paths,
which becomes small for the real-world networks when $\langle k \rangle$ is 1--10.
For these small $\langle k\rangle$, $S$ depends crucially
on the global topological organization: it is $\log_22=1$ for a one-dimensional string,
$\log_2N$ for a star, but of order $N/4$ for a stringy
structure with many separated branches of length 1 (see Fig.\ \ref{fig3}).
The increase of the average shortest path length $\langle l \rangle$, plotted as a dashed line,
indicates that the stringy structure dominates in the ensemble of random networks with low $\langle k \rangle$.

\begin{figure}[htbp]

\includegraphics[width=\columnwidth]{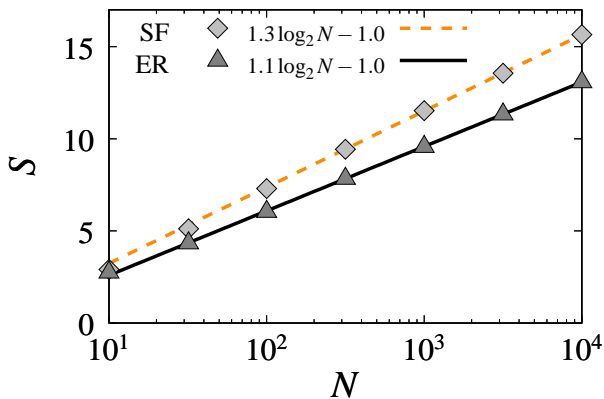}
\caption[] {(Color online) $S$ as function of system size $N$ for
fully connected random network topologies with fixed average degree
$\langle k \rangle = 5$.
ER refers to Erd{\H o}s-Rényi random networks and
SF to random scale-free networks with degree
distribution $\propto 1/(k_0+k)^{2.4}$ with
$k_0$ adjusted for every network size so that $\langle k \rangle$ is
kept fixed. A star network with one node connected to the
remaining nodes in the network and the remaining links randomly distributed scales
asymptotically as $S = \log_2N +1$ as in principle every shortest path goes 
through the hub with the cost $\log_2(N-1)$.}
\label{fig4}
\end{figure}

In Fig.\ \ref{fig4} we demonstrate that $s=S/\log_2(N)$
is nearly a size-independent way to compare
networks of different size with each other \cite{city}. Thus
this quantity is an invariant for any given
type of network topology,
whether it is dominated by a single hub (star), whether it is
scale-free (SF), or whether it is of Erd{\H o}s-Rényi (ER) type.
In all cases we compare networks with the same average
degree and find that $s$ nicely differentiates between 
different types of networks with
a given amount of links between the nodes.
The asymptotic logarithmic scaling can 
be understood by the logarithmic increase in average shortest path length
for Erd{\H o}s-Rényi networks, $\langle l_{sp} \rangle \propto \log N$ \cite{holyst}, and constant 
cost at every node ($\log_2 \langle k \rangle$).
For scale-free networks it is a little deeper, but simple for the extreme cases.
For $\gamma = 2$ the average shortest path length is constant and the
size of the largest hub in the network scales linearly with the system size \cite{holyst}.
As almost all shortest paths will go through this ``superhub'' as in a star network,
the search information is proportional
to $\log_2 N$. For $\gamma > 3$ the average shortest path length scales as $\log N$
and the largest hub is finite, similar to Erd{\H o}s-Rényi networks \cite{cohen}.

\begin{figure}[htbp]
\includegraphics[width=\columnwidth]{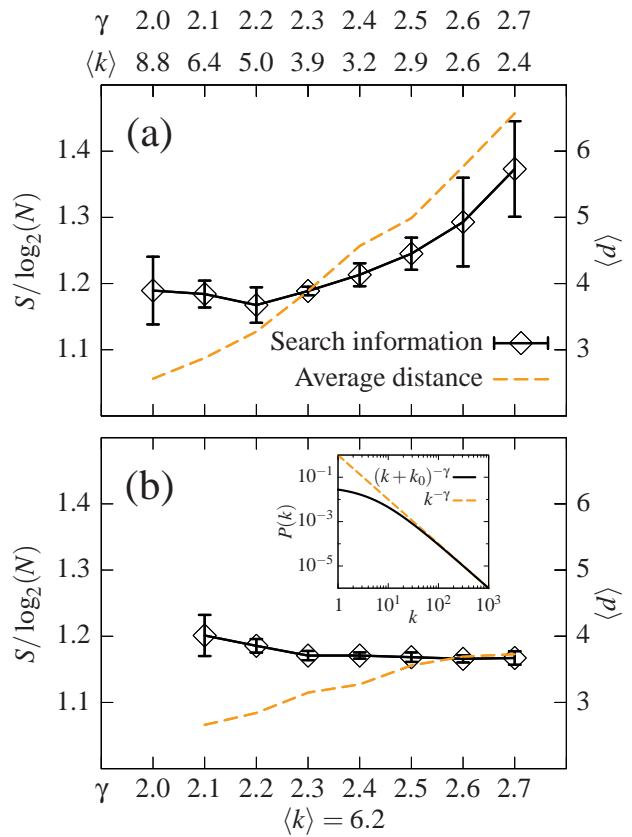}
\caption[] {(Color online) $S$ as a function of the exponent $\gamma$ for random scale-free
networks with degree distribution $P(k)\propto 1/k^{\gamma}$ in (a).
Varying $\gamma$ implies varying average degree $\langle k \rangle$
according to the second $x$-label row. An increased $\gamma$ also
implies a decreased frequency of large hubs and a lengthened 
average shortest path between nodes. To separate the effects we in (b)
study the distribution $P(k)\propto 1/(k_0+k)^{\gamma}$ 
with $k_0$ set to keep the $\langle k \rangle$ fixed, as in the insert. Here we see that
even though the average shortest path increases, $S$ decreases slightly
due to a decreased frequency of hubs. However, compared to (a) the
average shortest path increases substantially less with increasing $\gamma$,
due to the constant $\langle k \rangle$. The increasing average shortest path
accordingly dominates over the decreasing frequency of hubs for $\gamma > 2.2$
in (a). The size of the networks is $N=10^4$ in both (a) and (b).
}
\label{fig6}
\end{figure}

From Fig.\ \ref{fig4} we also notice
that scale-free networks have the largest $S$,
at least as long as we consider
a random organization of the topology. This is because
nodes with large values of $k_i$ also have large $b_i$,
and therefore contributes relatively more to
the overall confusion according to Eq.\ (\ref{simple}).
This fact is explored more in Fig.\ \ref{fig6}
where we show the variation of $S/\log_2(N)$ as function
of degree distribution quantified by $\gamma$.
At low $\gamma\sim 2$, where effectively
a scale-free network behaves very similar to
a star network, the largest
hubs tend to be
connected to a major fraction of the system.
A typical path therefore passes through
a major hub of degree $k \propto N$ and maybe one more
node as indicated by the average shortest path length.
For larger $\gamma$ the high cost of passing nodes with
$k \propto N$ disappears, but the total average cost
nevertheless increases since the path length increases 
rapidly. In Fig.\ \ref{fig6}(b) the average degree is 
kept constant by adjusting $k_0$ in the degree distribution
$P(k) \propto (k_0+k)^{-\gamma}$. This weakens the increase
in average path length as $\gamma$ increases and
$S$ instead slowly decreases because the probability for
having very large hubs decreases.
  
\begin{figure}[htbp]

\includegraphics[width=\columnwidth]{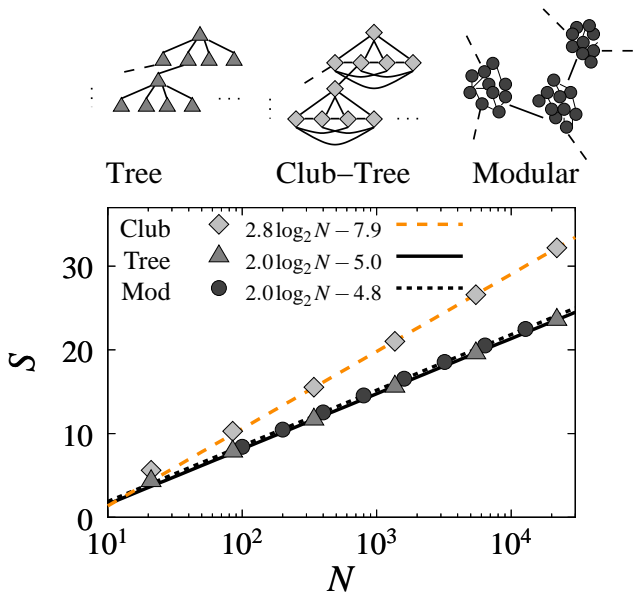}
\caption[] {(Color online) $S$ versus $N$ for tree, club-tree
and modular networks. Both trees have a branching ratio $d$ of 4, as seen in the illustration above.
The modular network consists of communities of ten nodes, each of them connected to 
five other nodes. Each community is in turn connected with three other communities.
}
\label{fig5}
\end{figure}

We now turn to networks with narrow degree distributions, but nonrandom
topologies and start with an illustrative calculation of $S$ for a tree hierarchy.
We obtain $S=2\log(N)-5$ numerically for trees of different branching ratios $d$ (Fig.\ \ref{fig5}),
which was corroborated analytically for a binary tree.
However, $S$ depends on addition of links to the tree, and in 
particular $S$ is larger for the club tree, as numerically demonstrated 
in Fig.\ \ref{fig5}.
In any case $S$ for trees is much larger than for random networks. 
The reason why trees are perceived as efficient is (1) that they are 
efficient seen from the top (e.g., data structures), 
and (2) trees are mostly associated not
to specific signaling, but rather to broadcasting of information, 
where everyone in a certain section is given the same information (e.g., military organization).
Even higher information cost has a regular network (every node connected to twice the dimension $d$ of the lattice) as
the shortest path length scales as $N^{1/d}$.
If the links of the regular network instead represents street segments between 
intersections in a square city like Manhattan (streets
and avenues mapped to nodes and intersections to links between the streets and avenues 
in a fully connected bipartite network \cite{city}), the result is
completely different.
Let $N$ streets be divided into $N/2$ north-south (NS) streets,
and $N/2$ east-west (EW) streets.
Going from any NS street to a particular EW street demands
information about which of the $N/2$ exits is correct. 
This information cost is $S(\mathrm{NS}\rightarrow \mathrm{EW})=\log_2(N/2)$.
To go from one NS street to another NS street means that any of
the $N/2$ EW streets can be chosen.
Each path is thus assigned a probability $(2/N)[1/(N/2-1)]$.
But there are in fact $N/2$ degenerate paths, and the 
total information cost for locating parallel
roads in this square city reduces to
\begin{equation}
\label{sq_city}
S(ns\rightarrow ns)\; =\;
- \log_2\left( \frac{N}{2} \frac{1}{N/2}
\frac{1}{N/2-1} \right)
=\log_2(N/2-1),
\end{equation}
reflecting the fact that it does not matter which of the EW roads one uses to reach the target road.
This places the fully connected bipartite network in the same class as the star network.

As an example of typical organization in social systems we also show the 
$N$ dependence of modular networks in Fig.\ \ref{fig5} \cite{girvan}. Again, $S$ is larger than in any
random network irrespective of the degree distribution.
We can therefore extend the previous statement 
that the value of $s=S/\log_2(N)$
is related to the global organization principle to include both
the degree distribution, Fig.\ \ref{fig4},
and the way the nodes are
positioned relatively to each other, Fig.\ \ref{fig5}.

\begin{figure}[htbp]
\includegraphics[width=\columnwidth]{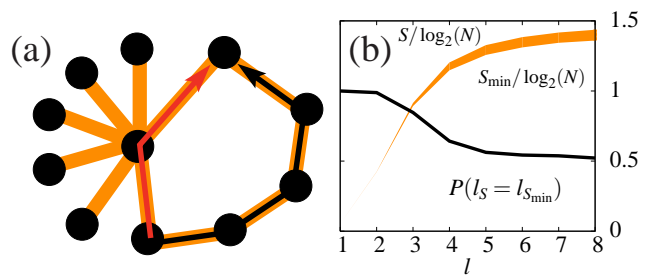}
\caption[] {(Color online) The minimum information path of length $l_{S_{min}}$ is the path between two nodes that regardless of distance has 
the smallest cost. In (a) it is clear that the right path is cheaper informationwise but longer in number of steps.
(b) shows the fraction of minimum information paths that are also shortest paths in a scale-free network with $N=10^4$ nodes and $\gamma=2.4$.
Although the overall fraction is as low as 0.62, $S_{min}$ is only
$5 \%$ smaller than $S$. As degeneracy is not considered in $S_{min}$ we compare with $S$ without degeneracy, $S_u$ as in Fig.\ \ref{fig2}(a).}
\label{fig6b}
\end{figure}

Again, all results are robust to the details in the formulation of $S$ and
very similar results would have been obtained if we instead had considered $S_u$
and excluded degeneracy or $S_d$ with a different weight of the degenerate paths.
To extend this we show in Fig.\ \ref{fig6b}(b)  the deviations between the search
information $S$ and a minimum search information $S_{min}$, where we take the minimum information
concept to the extreme and look for the path \emph{regardless of length} that has the
smallest information cost. This would typically be a path that avoids hubs. In Fig.\ \ref{fig6b} it is obvious
that the right choice is cheaper informationwise even though the path is longer.
Intuitively the number of shortest information paths that also are shortest paths will decay 
as the paths get longer and longer. This is confirmed in Fig.\ \ref{fig6b}(b) for
a scale-free network of size $N=10^4$ with $\gamma=2.4$. Nevertheless, the difference 
from the shortest information path is small. This observation is valid in
case of logarithmic information cost at every node, as in the present case.
If the cost instead was linear as in Fig.\ \ref{fig1b}(c), the difference would be substantial
as hubs would repel the minimum information paths much more.

\section{Node organization}
\begin{figure}[htbp]
\includegraphics[width=\columnwidth]{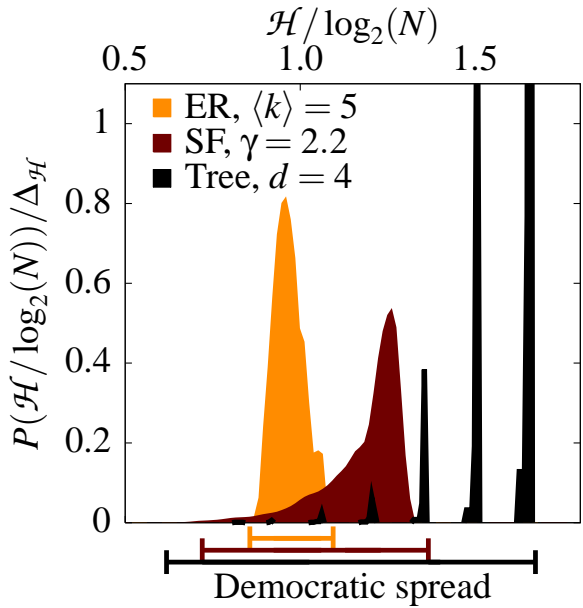}
\caption[] {(Color online) Distribution of hide $\mathcal{H}$ for nodes on various types of networks (low $\mathcal{H}$ means high visibility).
We see that the Erd{\H o}s-Rényi (ER) network is quite homogeneous, the scale-free (SF) network has a wider distribution,
whereas the tree hierarchy is by far the least democratic in distributing
the ability of different nodes to communicate.
The democratic spread plotted below the box roughly estimates 
the division of communication in the network.
The number of nodes is $N=10^4$ for all networks.
}
\label{fig7}
\end{figure}

We have until now presented a tool to characterize networks 
on the global level and quantified networks as being easy or difficult to navigate 
or search on average. 
We now turn to the effect the organization of 
networks has on the individual nodes.
The specific communication approach opens up a natural way
to characterize the different networks in terms of
their ability to distribute communication options among their nodes.
We therefore define the hide
\begin{equation}\label{hide}
\mathcal{H}_t=\sum_s S(s \to t)/N
\end{equation} 
as the average number of bits a walker needs to walk directly from a random node
in the network to the target node $t$ \cite{pnas}.
The different values of hide reflect to what degree the nodes are visible.
Low hide $\mathcal{H}$, or low average information cost to find the node, represents
high visibility.
This is illustrated in Fig.\ \ref{fig7}, where  in agreement with
intuition we find that Erd{\H o}s-Rényi networks are by far the most
democratic, whereas scale-free and especially tree hierarchies are
hugely elitist. In particular the tree hierarchy has localized all
communication (low $\mathcal{H}$ means high visibility and thus ability to
receive information) to the top nodes. In Fig.\ \ref{fig7} we plot the democratic spread
as the difference between the most and the least visible node in the network as an 
illustrative estimate of the distribution of communication in the network.

\begin{figure}[htbp]
\includegraphics[width=\columnwidth]{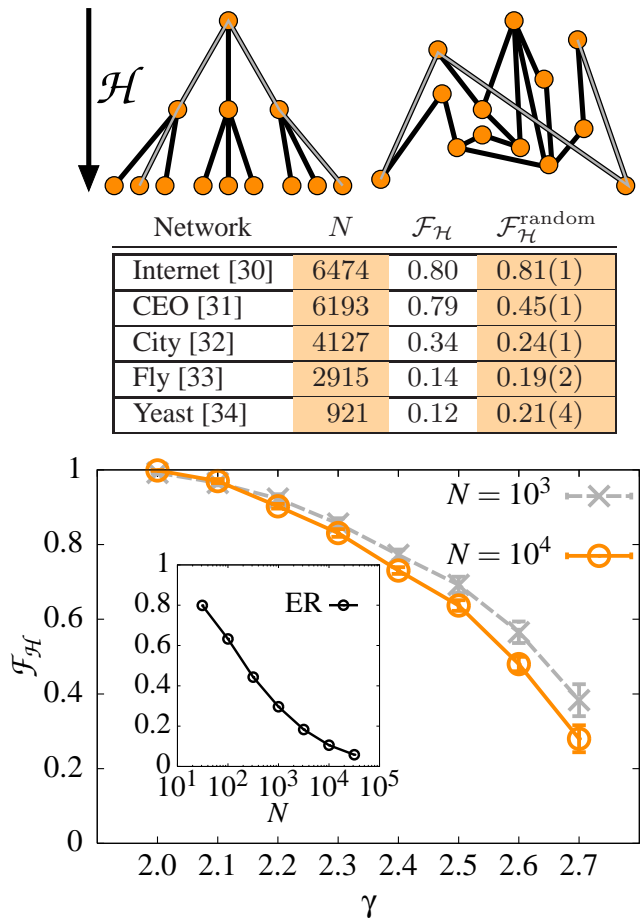}
\caption[] {(Color online) The hierarchical features of scale-free and Erd{\H o}s-Rényi 
networks with hide $\mathcal{H}$ representing the importance of a node.
The nodes in the networks in the top of the figure are arranged in hierarchical order.
The intuitively hierarchical order of the tree is reproduced with the $\mathcal{H}$ ordering 
and all shortest paths are hierarchical, $\mathcal{F}_{\mathcal{H}} = 1$. The network to the right 
has the opposite property. A high ratio of the shortest paths are not hierarchical since 
a typical path repeatedly goes up and down in the $\mathcal{H}$ hierarchy.
The table shows a number of real-world networks and their $\mathcal{F}_{\mathcal{H}}$ together with
the corresponding value in randomized networks with the same degree sequence \cite{maslov2002,maslovInternet}.
The biological networks are randomized so that both bait and prey degree of all proteins
are preserved. The plot in the bottom of the figure shows the behavior of scale-free and Erd{\H o}s-Rényi networks with respect
to $\mathcal{F}_{\mathcal{H}}$. Scale-free networks are $\gamma$ dependent whereas Erd{\H o}s-Rényi networks are size dependent.
}
\label{fig8}
\end{figure}

The different degrees of hide information of the various nodes
effectively rank the nodes, and thereby suggest a self-consistent
measure of a hierarchy based on visibility.
At the same time the hide $\mathcal{H}$ captures both the hierarchy in the usual terms
of trees, as in military structures, and the intrinsic hierarchical nature of 
topological hierarchies for scale-free networks \cite{hierarchy} as in the Internet \cite{maslovInternet}.
A highly ranked node is close to the top in a tree. The corresponding node
in a topological hierarchy is a highly connected node. In the Internet, for example,
the highly connected nodes play the roles of intermediate
nodes on typical paths between nodes further down in the hierarchy, just like top nodes in a tree. 
In analogy with
\cite{gao,hierarchy} we define a path from $s$ to $t$ to be hierarchical
if it defines a common boss for $s$ and $t$. That is, the path has
first to decrease monotonically in $\mathcal{H}_j$ to more and more visible nodes, until a minimum, and
thereafter increase monotonically in $\mathcal{H}_j$ until the target node
$j=t$ is reached. We allow the path to pass between nodes with the same
value of $\mathcal{H}_j$, and we consider paths that only increase or only
decrease as hierarchical. Given $\mathcal{H}_j$ for each node $j\in[1,N]$ in
a network we quantify the network's degree of information hierarchy $\mathcal{F}_\mathcal{H}$  
by the fraction of shortest paths between nodes in the
network which are also hierarchical paths:
\begin{equation}
\mathcal{F}_\mathcal{H} = \frac{(\mathrm{number\: of\: hierarchical\: shortest\: paths})}
{N(N-1)},
\end{equation}
where the denominator counts the total number of shortest paths between nodes in the network. 
In case of degenerate shortest paths, each path contributes to
$\mathcal{F}_\mathcal{H}$ by a weight given by its contribution to the traffic
betweenness. In accordance with intuition we find that $\mathcal{F}_\mathcal{H}$ decays
with system size for random Erd{\H o}s-Rényi networks, as 
shortest paths get longer. The decay is plotted in the inset of Fig.\ \ref{fig8}. 
$\mathcal{F}_\mathcal{H} =1$ for both hierarchies and club hierarchies whereas $\mathcal{F}_\mathcal{H}$ for
random scale-free networks depends on degree distribution. Figure \ref{fig8} shows how the information hierarchy
varies with degree distribution for pure random scale-free networks parametrized by
$P(k)\propto 1/k^{\gamma}$. As $\gamma$ increases from $2$ the network goes from being a 
complete information hierarchy with $\mathcal{F}_\mathcal{H} =1$ toward $0$ when $\gamma$ 
approaches $3$, the average degree approaches $2$, and shortest paths become long.
For real-world networks the overall observation is that biological networks are antihierarchical with respect to
$\mathcal{F}_\mathcal{H}$, while social and communication networks tend to be hierarchical (see table in Fig.\ \ref{fig8}).
The Internet is a network of autonomous systems \cite{internet} that 
in this data set consists of 6474 nodes and 12 572 links and its degree 
distribution is scale-free with $P(k)\propto 1/k^{2.1}$.
In the CEO network (6193 nodes and 43 074 links),
chief executive officers are connected by links if
they sit on the same board \cite{ceo}.
The city network is constructed by mapping 4127 streets to nodes and
5565 intersections to links between the nodes in the Swedish city of Stockholm \cite{city,teleadress}.
Fly is the protein interaction network
in \emph{Drosophilia melanogaster} 
detected by the two-hybrid experiment \cite{giot},
and yeast refers to the similar network
in \emph{Saccharomyces cerevisiae} \cite{Uetz2000}.

Overall, for scale-free networks, the information hierarchy $\mathcal{F}_\mathcal{H}$ 
quantitatively follows the topological hierarchy $\mathcal{F}$ presented in \cite{hierarchy}.
Thus networks with maximal (minimal) topological
hierarchy $\mathcal{F}$ \cite{hierarchy}, also have large
(small) $\mathcal{F}_\mathcal{H}$. But it is important that the information hierarchy allows for
a natural generalization to non-scale-free networks, and is therefore 
a unified definition of hierarchical organization with the most
visible node in the top. A less powerful ranking is the betweenness \cite{between} as the betweenness
is sensitive to links that shortcut important nodes. By adding links between the
children of a top node as in the club tree in Fig.\ \ref{fig5},
the ranking changes completely as the betweenness for the top node in principle would be zero,
whereas its position at the top would still be reflected by the hide ranking. 

\section{Conclusion}
Networks are a natural way to visualize the limited information access 
experienced by individual parts of the overall system.
In the present paper we have explored topologies of a number of model 
networks in terms of their ability to facilitate peer-to-peer communication.
The ability to transmit specific signals is quantified in terms
of the difficulty in navigating the networks, quantified by the search information $S$.
As an overall lesson we have found that the inequality
\begin{align}
S(\mathrm{real\;world}) > S(\mathrm{random,fixed\;degree}) > S(\mathrm{ER}),
\end{align}
is valid for all investigated real-world networks \cite{pnas,horizon,city,bit}.
Here $S(\mathrm{random,fixed\;degree})$
represents randomized networks with exactly the same
degree distribution as the investigated real-world network,
whereas ER (Erd{\H o}s-Rényi) networks only have the same total number of nodes and 
links as the real-world network.
The above inequality is in particular associated with cases where the cost of passing a node is 
proportional to $\log(k)$, 
but it is also true for the higher local information cost proportional to $k$,
where $k$ is the degree of the node.
As $S$ represents an average of the contribution from any node to any other node,
the major contribution to $S$ comes from pairs of nodes that are separated 
by large distances. The fact that $S$ in realistic networks is relatively large teaches us
that the topology of real-world networks disfavors distant specific communication \cite{horizon,bit}.
Topologically, large $S$ was found in a number of model networks, 
with modular or hierarchical features with highly connected nodes deliberately positioned
``between'' other nodes, hinting that a large search information $S$ is associated
not only with broad degree distributions, but also with well known organizational
features of social and biological systems.

The peer-to-peer search information $S(s\rightarrow t)$ opens the possibility for
a detailed measure of the relative ``importance'' of nodes in a given network.
In fact, measuring visibility of a node $t$ in terms of how well hidden the node
is from the rest of the network as in Eq.\ (\ref{hide}),
we have shown how networks can be ranked in terms of a generalized hierarchy measure.
The measure captures both the hierarchy in the usual terms of trees shown in Fig.\ \ref{fig7}
and at the same time also the intrinsic topological hierarchical nature of 
scale-free networks.
Thus, this generalized hierarchy measure defines scale-free networks with degree
distribution with exponent close to $\gamma=2$ to be hierarchical, whereas 
narrower distributions will be antihierarchical unless they are deliberately 
organized in a treelike structure.

Overall, the different ways of organizing 
networks can be recast according to their ability or inability to transmit 
specific messages across the networks.
The presented search information $S$
provides a useful measure of this key functional role that
is reflected in the topology of many real-world networks.
\bigskip

\section*{ACKNOWLEDGMENTS}
We acknowledge the support of Swedish Research Council through
Grants No.\ 621 2003 6290 and No.\ 629 2002 6258 and the center 
Models of Life supported by Danmarks Grundforskninsfond.


\vfill\eject

\end{document}